\documentclass[
    ,final            % use final for the camera ready runs
  ]
  {aipproc}

\layoutstyle{6x9}

\begin{document}

\title{Spin in Hadron Reactions}


\author{Christine A. Aidala}{
  address={University of Massachusetts, Amherst, MA 01003-9337, U.S.}
}

\begin{abstract}
 The Relativistic Heavy Ion Collider (RHIC) has brought the study of spin effects in hadronic collisions to a new energy regime.  In conjunction with other experiments at facilities around the world, much can be learned from the high-energy polarized proton collisions RHIC provides, allowing the collider to serve as a powerful tool to continue to understand the rich subtleties and surprises of spin effects in QCD, some of which were originally discovered more than three decades ago.
\end{abstract}

\maketitle

%%%%%%%%%%%%%%%%%%%%%%%%%%%%%%%%%%%%%%%%%%%%
%% MAINMATTER
%%%%%%%%%%%%%%%%%%%%%%%%%%%%%%%%%%%%%%%%%%%%

\section{Introduction}

Since the startling discovery of hyperon polarization in collisions of an unpolarized proton beam on a beryllium target in 1976, hadronic interactions have played an essential role in exploring spin effects in quantum chromodynamics (QCD).  The Relativistic Heavy Ion Collider (RHIC), inaugurated in 2001 as a high-energy polarized proton collider, has taken the stage as the flagship facility for the exploration of spin in hadron reactions in recent years.  Having achieved polarized proton collisions at center-of-mass energies as high as 200~GeV with approximately 60\% polarization, and with polarized proton collisions at 500~GeV planned for 2009, RHIC offers a wealth of opportunities.   The collider has already made a number of contributions toward unraveling the mysteries of spin effects in QCD, which can broadly be placed into three categories: hyperon polarization, the transverse spin structure of the nucleon and transverse-momentum-dependent parton distributions, and the helicity structure of the nucleon.

\section{Hyperon Polarization}

More than three decades ago, an experiment at Fermilab was performed in which a 300-GeV beam of unpolarized protons was fired at a beryllium target, and lambda hyperons were measured in the final state.  The self-analyzing decay of the lambda into a proton and a pion allowed determination of the polarization of the produced particles, and remarkably, transverse polarization of up to nearly 20\% was observed, increasing with the transverse momentum ($p_T$) of the lambda up to the highest measured value of 1.6~GeV/$c$ \cite{Bunce:1976yb}.  More than 30 years after this initial discovery, and despite plentiful experimental confirmations of the effect, the mechanism by which the hyperons acquire polarization remains very poorly understood.

\begin{figure}
  \includegraphics[height=.18\textheight]{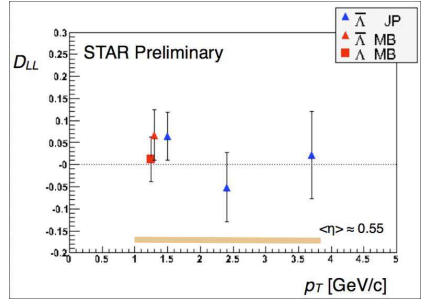}
  \caption{The spin transfer from a longitudinally polarized proton to a longitudinally polarized lambda or antilambda produced in proton-proton collisions at $\sqrt{s}=200$~GeV, measured by STAR.  "JP" and "MB" denote data taken with the STAR jet patch and minimum-bias triggers, respectively.}
  \label{fig:starSpinTransfer}
\end{figure}

\begin{figure}
  \includegraphics[height=.18\textheight]{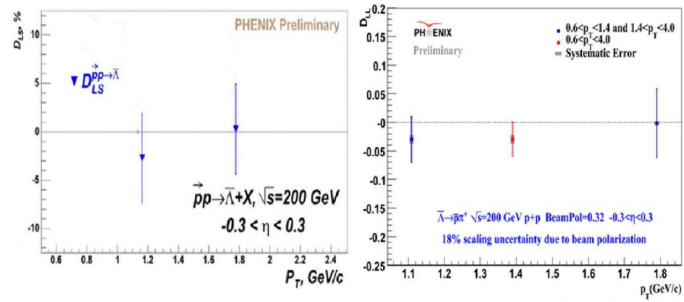}
  \caption{The spin transfer from a longitudinally polarized proton to a transversely (left) or longitudinally (right) polarized antilambda produced in proton-proton collisions at $\sqrt{s}=200$~GeV, measured by PHENIX.}
  \label{fig:phenixSpinTransfer}
\end{figure}

With the availability of polarized beams, the polarization \emph{transfer} to final-state hyperons can instead be measured, providing new pieces to help resolve the puzzle.  This was done in the mid-1990s by the E704 experiment at Fermilab, and a significant correlation between the transverse spin of the initial-state proton and the final-state lambda of up to $\sim 30$\% was observed, rising again with $p_T$ \cite{Bravar:1997fb}.  The polarized proton beams at RHIC allow similar measurements to be made.  In Figs.~\ref{fig:starSpinTransfer}-\ref{fig:phenixSpinTransfer} are shown spin-transfer measurements from initial-state, longitudinally polarized protons to final-state, transversely or longitudinally polarized (anti-)lambdas.  The polarization transfer in both cases is consistent with zero for all measured $p_T$.  Spin-transfer results with transverse polarization in the initial state are expected from RHIC in the future and will be important to compare to the E704 result performed at a center-of-mass energy one order of magnitude lower.  While hyperon polarization remains poorly understood, new data combined with present work on transverse spin phenomena may eventually shed some light on the matter.

\section{Transverse Spin Phenomena}

The study of transverse spin phenomena has witnessed a veritable explosion of activity over approximately the past seven years.  For more detailed perspectives of the field than given here, see e.g. \cite{Anselmino:2009tp} and \cite{Makins:2009tp}. Surprisingly large single-spin asymmetries (SSA's) of magnitudes as large as 40-50\% were initially observed in collisions of transversely polarized protons with other hadrons at the ZGS at Argonne National Laboratory \cite{Klem:1976ui,Dragoset:1978gg} as well as the PS at CERN \cite{Antille:1980th} in the late 1970s and early 1980s, for center-of-mass energies between 5 and 10~GeV.  It was these early hadronic data as well as preliminary hadronic data from Serpukhov, Protvino that inspired the groundbreaking paper by Dennis Sivers in 1989 \cite{Sivers:1989cc}, which proposed that these large asymmetries could be generated by a transverse-momentum-dependent partonic distribution (TMD), in which the transverse momentum of the partons is correlated with the spin of the proton.

\begin{figure}
  \includegraphics[height=0.25\textheight]{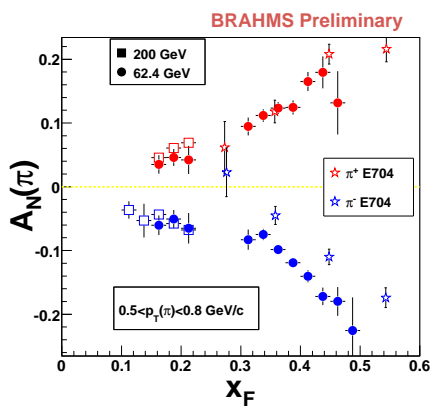}
\vspace{-7 mm}
\caption{Charged pion asymmetries measured at 200 and 62.4~GeV by the BRAHMS experiment and at 19.4~GeV by the E704 experiment, shown for overlapping kinematic ranges.  }
\label{fig:threeEnergies}
\end{figure}

Subsequent experimental data over a wide range of energies indicate that similar effects exist for particle production at center-of-mass energies of tens of GeV and even as high as 200~GeV at RHIC. In Fig.~\ref{fig:threeEnergies} are shown the transverse SSA's for charged pions measured at 19.4 \cite{Bravar:1997fb}, 62.4 \cite{Arsene:2008mi}, and 200~GeV.  The higher-energy data allow perturbative QCD (pQCD) to be used in attempting to interpret the results.  While much remains to be learned, exciting progress has been made in understanding the observed transverse single-spin asymmetries at high energies, and the wealth of experimental data now becoming available from RHIC and other experiments is helping to drive the field forward.

The complex task of understanding measurements performed in hadronic collisions is now being facilitated as relevant quantities are being studied in semi-inclusive deep-inelastic scattering (SIDIS) and $e^+e^-$ annihilation.  A definitively non-zero transverse-momentum-dependent Collins fragmentation function (FF) measured in $e^+e^-$ annihilation was published by the BELLE experiment in 2005 \cite{Abe:2005zx}.  The availability of the Collins FF makes it possible to extract the transversity distribution from asymmetry measurements in SIDIS and proton-proton collisions, and first extractions of transversity have been published \cite{Anselmino:2007fs}.  First extractions of the Sivers TMD from SIDIS data have also been released \cite{Anselmino:2008sga,Arnold:2008ap}.  As constraints start to be provided on transversity and the various transverse-momentum-dependent distribution and fragmentation functions, more can in turn be learned from hadronic collision data.

\begin{figure}
  \includegraphics[width=0.45\textwidth,height=0.18\textheight]{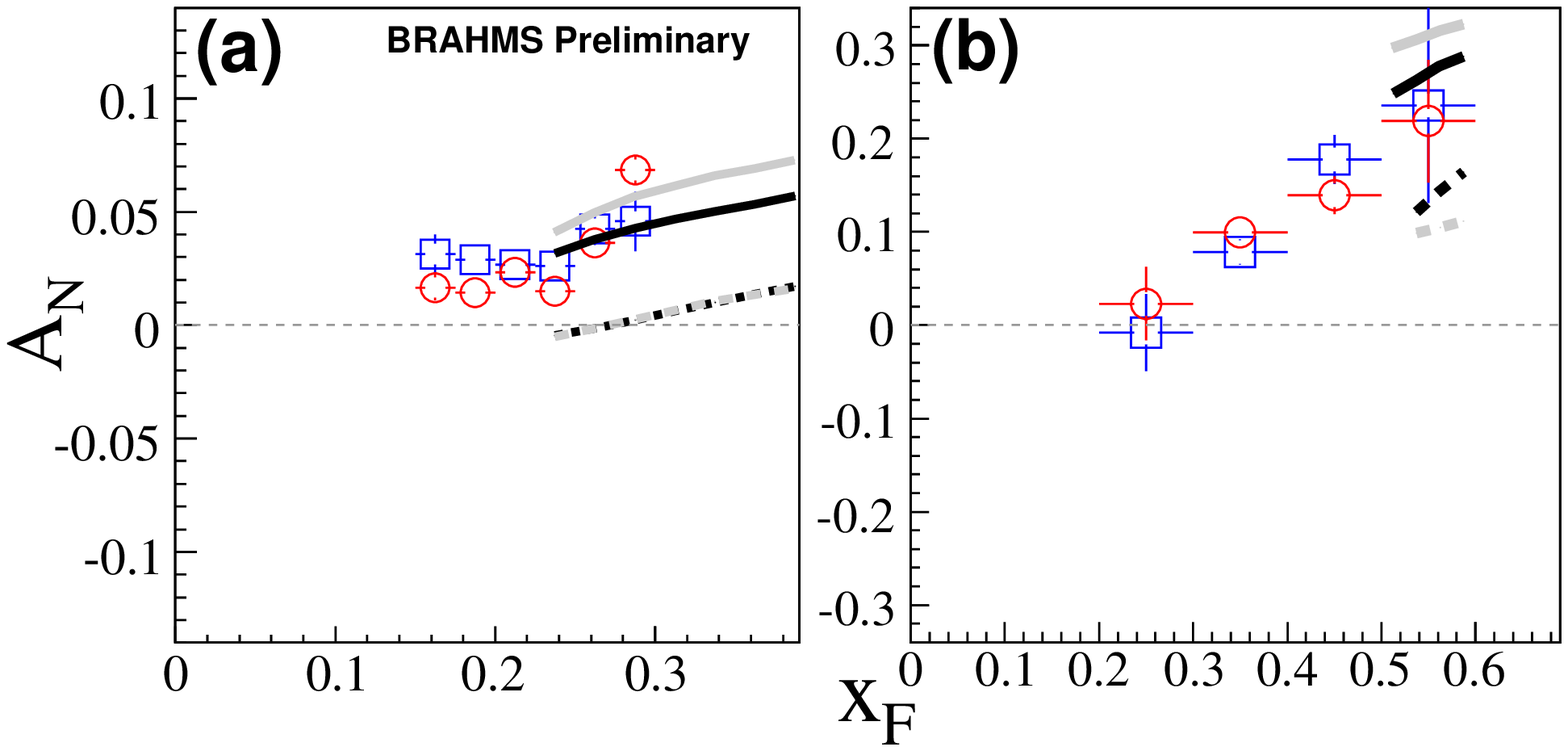}
\hspace{0.03\textwidth}
\includegraphics[width=0.39\textwidth,height=0.18\textheight]{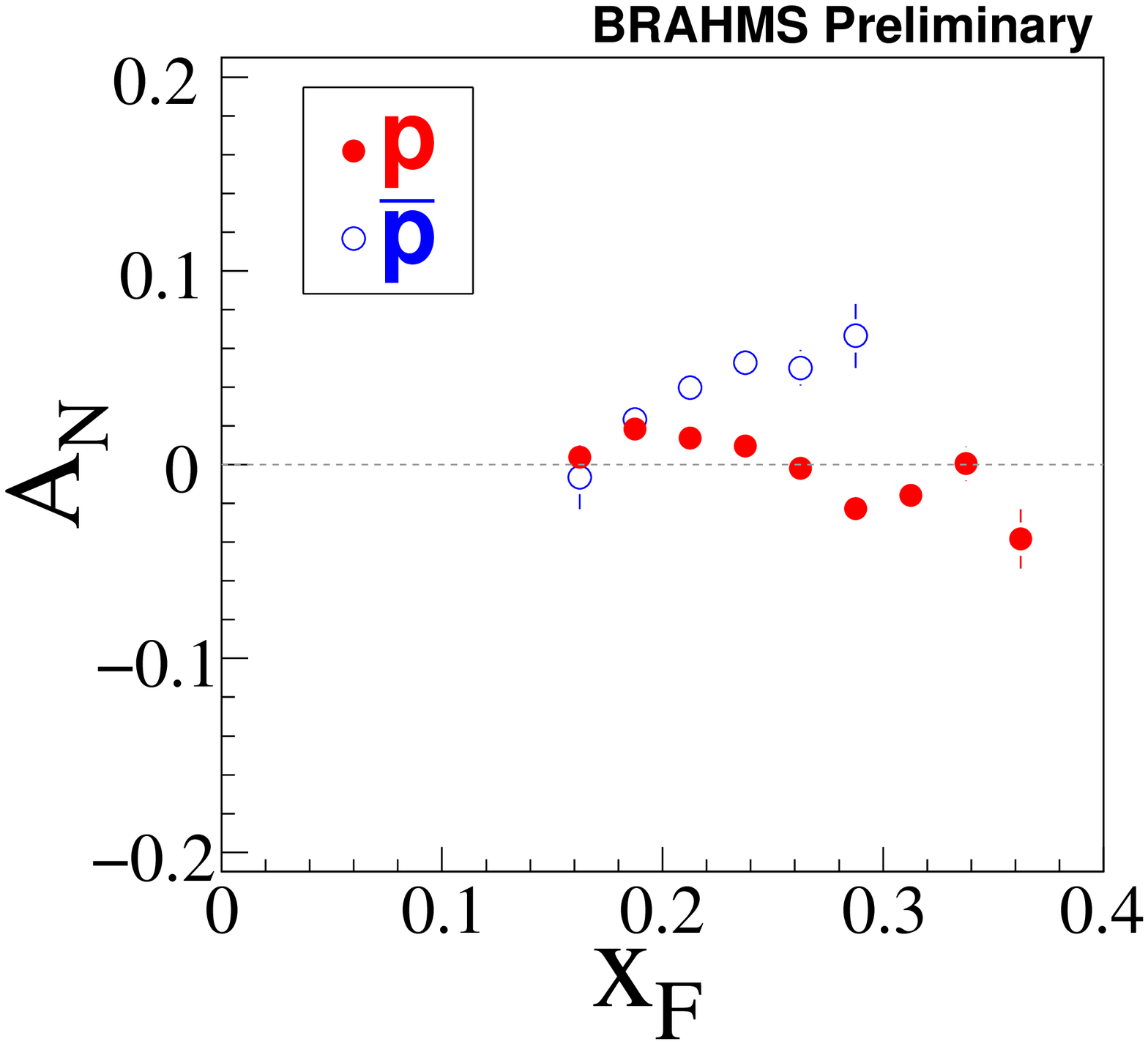}
\caption{Transverse SSA's measured by BRAHMS.  Kaons at 200~GeV (a) and 62.4~GeV \cite{Arsene:2008mi} (b).  Positive (negative) kaons are indicated by open circles (squares). Solid (dashed) curves are for positive (negative) kaon initial-state twist-three calculations with (black) and without (grey) sea-quark contributions.  Protons and antiprotons at 200~GeV (right).}
\label{fig:kaons}
\end{figure}

The transversely polarized proton-proton data from RHIC have provided several surprises to the field, all of which remain unexplained.  In 2002, a transverse SSA of approximately -10\% was observed in the production of forward neutrons for $p+p$ collisions at 200~GeV \cite{Bazilevsky:2006vd}, and a significant asymmetry has been seen to persist in collisions up to 410~GeV.  As can be seen in Fig.~\ref{fig:kaons} (a, b), at both 200 and 62.4~GeV, the SSA's in forward charged kaon production are very similar for both kaon species.  With the observed forward pion asymmetries understood as largely a valence quark effect, calculations of the SSA for negative kaons consistently underpredict the data.  Even more startling is the significant non-zero antiproton asymmetry discovered by the BRAHMS experiment at 200~GeV, shown in Fig.~\ref{fig:kaons} (right), while a SSA consistent with zero was observed for protons at both 200 and 62.4~GeV.  Unfortunately no measurement of antiprotons is available at 62.4~GeV.

\begin{figure}
  \includegraphics[width=0.4\textwidth,height=0.2\textheight]{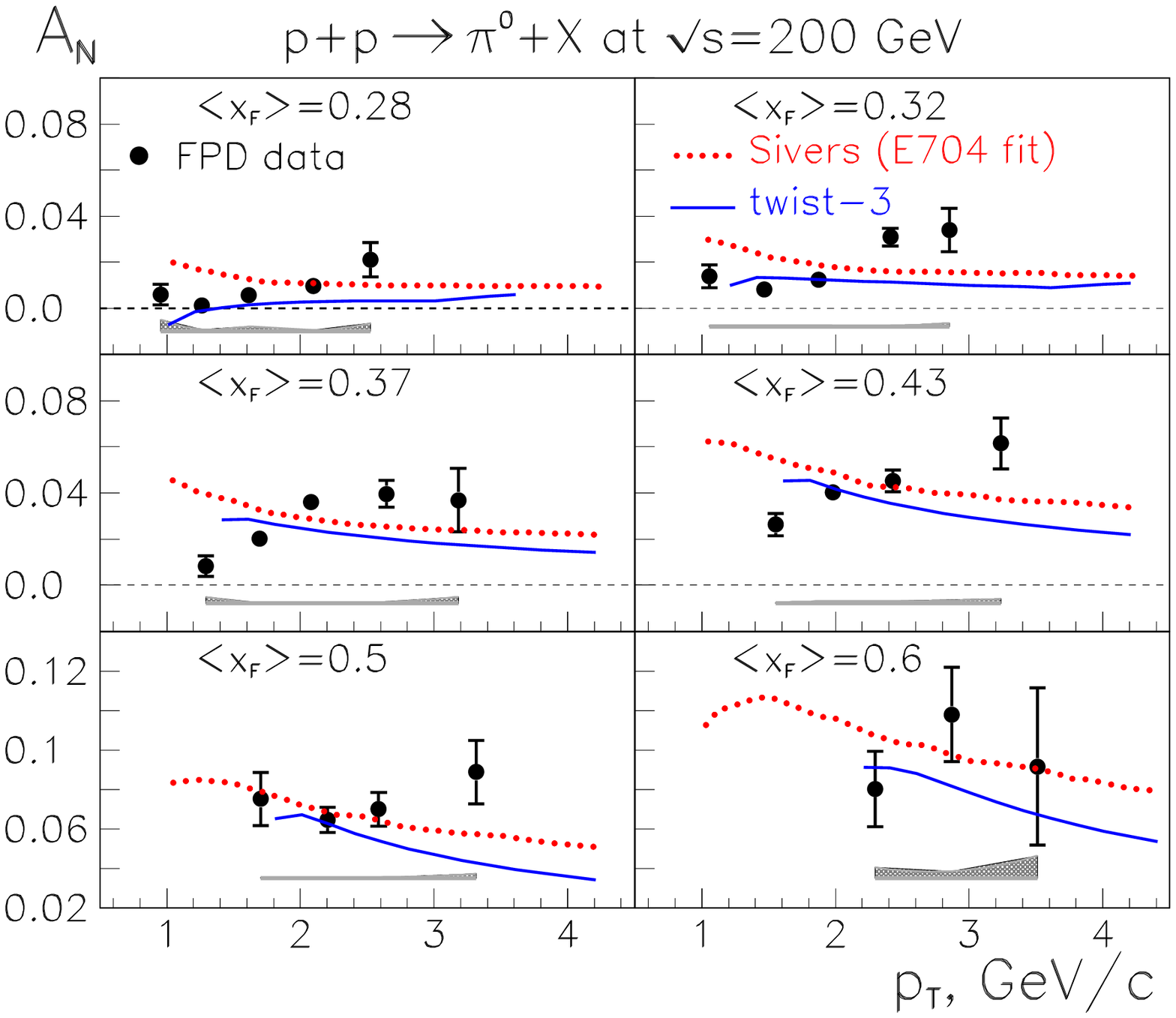}
\hspace{0.03\textwidth}
\includegraphics[width=0.4\textwidth,height=0.2\textheight]{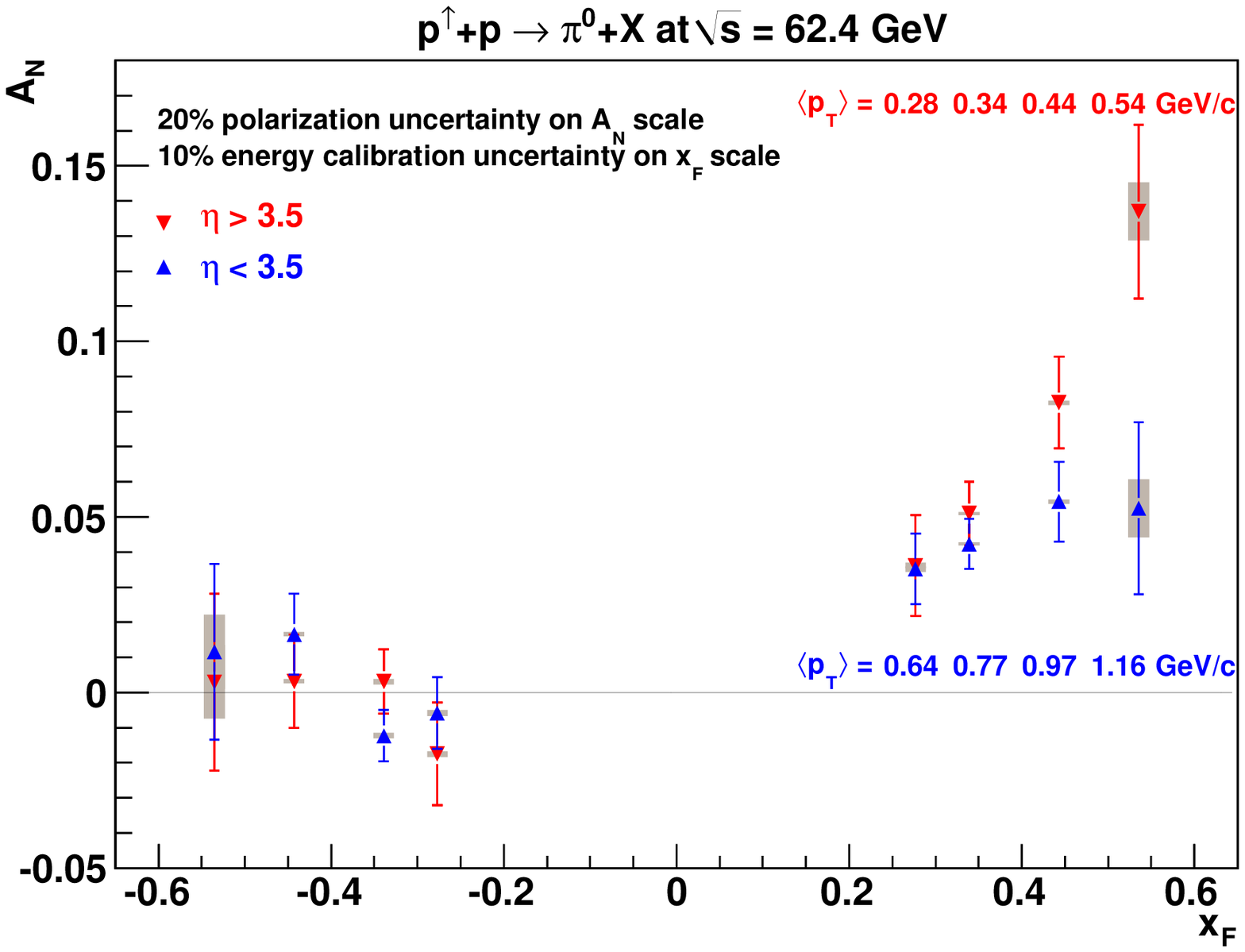}
  \caption{Left: $p_T$ dependence of the transverse single-spin asymmetry of neutral pions produced at $\sqrt{s}=200$~GeV, measured by STAR \cite{Abelev:2008qb}.  Right: Transverse single-spin asymmetry of neutral pions produced at $\sqrt{s}=62.4$~GeV for two different pseudorapidity ranges, measured by PHENIX \cite{Chiu:2007zy}.}
  \label{fig:starPi0AN}
\end{figure}

Theoretical work continues to attempt to unravel these mysteries, while the experimental data from RHIC become increasingly precise.  For inclusive pion measurements, enough statistics are now available to allow studies of the asymmetry dependence on $p_T$ and rapidity in addition to $x_F$, as shown for example in Fig.~\ref{fig:starPi0AN}.  As can be seen, the $p_T$ dependence of the data shown disagrees with existing models.  Close examination of the behavior of the SSA's as a function of various kinematic variables will be essential to constrain phenomenological models of the underlying physics generating the large asymmetries.

\section{Helicity Structure of the Nucleon}

Since the so-called "proton spin crisis" was discovered by the EMC experiment more than 20 years ago \cite{Ashman:1987hv}, the quest has been on to understand the contributions of the spins of the gluons and the sea quarks to the total spin of the nucleon. RHIC, capable of polarized proton collisions for $50 < \sqrt{s} < 500$~GeV, is in a unique position to study both the gluons as well as the sea quarks.  Performing hadronic collisions rather than deep-inelastic scattering, gluons contribute to the measured asymmetries at leading order.  With polarized proton collisions at 500~GeV planned to start in 2009, the parity-violating longitudinal single-spin asymmetry in $W$ boson production will allow flavor separation of the spin contributions of the sea quarks without reliance on fragmentation functions.

\begin{figure}
  \includegraphics[height=0.24\textheight]{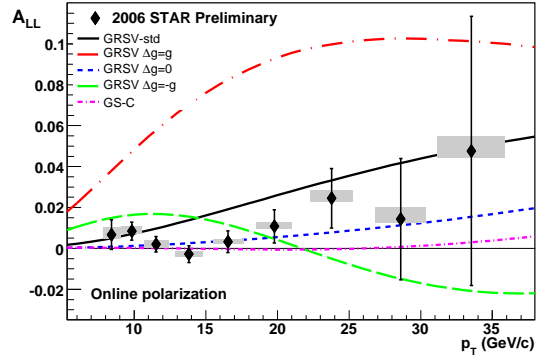}
\caption{The double-helicity asymmetry in inclusive jet production in $p+p$ collisions at $\sqrt{s}=200$~GeV \cite{Sarsour:2009tp}, measured by STAR, compared to predictions based on several assumptions for $\Delta G$ within the GRSV parameterization \cite{Gluck:2000dy}.}
\label{fig:jetAsym}
\end{figure}

\begin{figure}
  \includegraphics[width=0.7\textwidth,height=0.24\textheight]{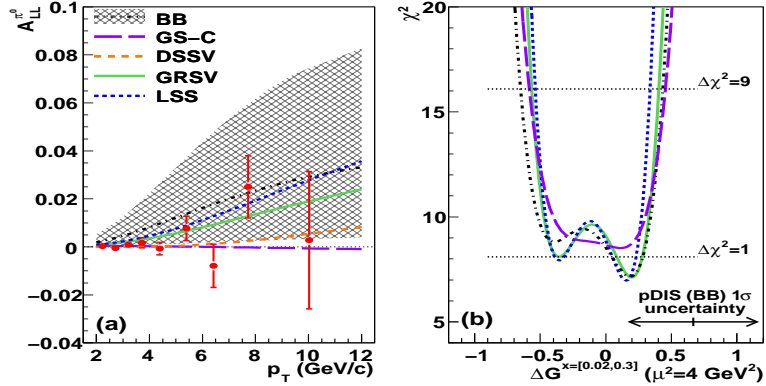}
\caption{The double-helicity asymmetry in inclusive neutral pion production in $p+p$ collisions at $\sqrt{s}=200$~GeV, measured by PHENIX, compared to expectations from several different parameterizations of the parton helicity distributions.  See \cite{Adare:2008px} for further details.}
\label{fig:pi0Asym}
\end{figure}

Several measurements have already been made at RHIC of double-helicity asymmetries for inclusive jet and hadron production, sensitive to $\Delta G$, the gluon spin contribution to the spin of the proton.  Jets represent the most abundant probe of $\Delta G$ available to STAR; PHENIX, with smaller acceptance and excellent electromagnetic calorimetry, has focused primarily on measurements of neutral pions.  The latest results from both experiments at $\sqrt{s}=200$~GeV can be seen in Figs.~\ref{fig:jetAsym} \cite{Sarsour:2009tp} and \ref{fig:pi0Asym} \cite{Adare:2008px}.  These measurements are sensitive only to a finite range in $x_{\rm gluon}$.  One way to extend the $x$ range probed is by varying the center-of-mass energy.  Measurements of neutral pions have already been made at 62.4~GeV \cite{Adare:2008qb}, and 500~GeV collisions will begin this year.

Precise determination of parton distribution functions requires many experimental measurements covering a broad kinematic range. The proton-proton collision data from RHIC have already helped to constrain $\Delta g(x)$ \cite{deFlorian:2008mr}, and further data will provide even stronger constraints over a wider kinematic range.

\section{Conclusion}
Hadronic reactions have a long history of seminal contributions to the study of spin effects in QCD.  In particular in conjunction with knowledge gained from simpler experimental systems, much can be learned from the more complex interactions between hadrons.  RHIC, as a high energy polarized proton collider, has opened up entirely new opportunities within the field; a wide variety of new results has already been released, and the program looks forward to many more years of data.

%%%%%%%%%%%%%%%%%%%%%%%%%%%%%%%%%%%%%%%%%%%%%%%%
%% BACKMATTER
%%%%%%%%%%%%%%%%%%%%%%%%%%%%%%%%%%%%%%%%%%%%%%%%



\bibliographystyle{aipprocl} % if natbib is missing

\bibliography{aidala_christine_spin2008_proceedings}

\IfFileExists{\jobname.bbl}{}
 {\typeout{}
  \typeout{******************************************}
  \typeout{** Please run "bibtex \jobname" to optain}
  \typeout{** the bibliography and then re-run LaTeX}
  \typeout{** twice to fix the references!}
  \typeout{******************************************}
  \typeout{}
 }

\end{document}